\documentclass[aps,twocolumn]{revtex4}
\def\beq{\begin{eqnarray}}
\def\eeq{\end{eqnarray}}
\begin{document}
\title{A comment on black hole state counting in loop quantum gravity}
\author{A. Ghosh}
\author{P. Mitra}
\affiliation{Saha Institute of Nuclear Physics,
1/AF Bidhannagar, Calcutta 700064}
\begin{abstract}
There are two ways of counting microscopic states of black holes in 
loop quantum gravity, one counting all
allowed spin network labels $j,m$ and the other involving only the labels $m$.
Counting states with $|m|=j$, as done in a recent Letter, does not follow either.
\end{abstract}
\pacs{04.70.Dy, 04.60.Pp}
\maketitle

Loop quantum gravity has yielded detailed counts of
microscopic quantum states corresponding to a black hole.
A start was made in
\cite{ash} in the direction of quantizing a black hole 
characterized by an
isolated horizon. The quantum states arise
when the cross sections of the horizon
are punctured by spin networks. The spin quantum numbers
$j,m$, which characterize the punctures, can label the quantum states. The entropy
is obtained by counting states 
that are consistent with a fixed area of the cross section
\cite{ash} and a total spin projection constraint.
An estimation of the entropy was carried out in \cite{meissner}
counting only $m$-labels (pure horizon states) -- see also \cite {gm2}. 
In \cite{gm}, the $j$-labels were also recognized as characterizing states.
Both approaches follow discussions in \cite{ash}.
Unlike these {\em approximate estimations}, \cite{cor} has used exact numerical methods,
counting $j,m$-labels as in \cite{gm}. Recently, \cite{agullo}
has also attempted exact calculations using some number theory. 
\cite{agullo} presents two
calculations, one of which counts $j,m$-labels, but the other counts only
states having $|m|=j$, {\em in a bid to follow} \cite{meissner}.
Unfortunately, as explained below, this prescription is {\em not} in general
equivalent to the rule of counting all horizon states or $m$-labels:
it was reached only {\em approximately} for large black holes \cite {meissner,gm2}. 
Consequently this counting of states in \cite{agullo} is invalid.

We use units such that $4\pi\gamma\ell_P^2=1$, where $\gamma$ is the 
so-called Barbero-Immirzi parameter involved in the quantization
and $\ell_P$ the Planck length. Setting the area $A$ of the
horizon equal to an eigenvalue of the area operator, we write
\beq 2\sum_{j,m}s_{j,m}\sqrt{j(j+1)}=A,\label{areacon}\eeq
where $s_{j,m}$ is the number of punctures carrying spin quantum numbers $j,m$ 
and obeying the {\em spin projection constraint}  
\beq \sum_{j,m}ms_{j,m}=0\;.\label{newc}\eeq

Consider for definiteness a small black hole with $A=4\sqrt{6}$.
This corresponds to 2 punctures each with $j=2$. Each puncture
in principle has 5 allowed values for $m$, but not all the 25
states obey (\ref{newc}), which is satisfied only if
$m_2=-m_1$, so that there are 5 states. These 5 states have
different $j,m$-labels and therefore the number of states
in the $j,m$-counting of \cite{gm} is 5. This is of course
what the $j,m$ calculation of \cite{agullo} yields. But in fact the
$j$-values of the two punctures being the same, the states
have different $m$-labels, so that the number of states in
the pure $m$-label counting envisaged in \cite{ash} is also
5. On the other hand, the number of states which the $|m|=j$ calculation
of \cite{agullo} gives is only 2, namely
the states with $m_2=-m_1=\pm 2$.

Consider next the situation $A=2\sqrt{2}+2\sqrt{6}$.
Here, there are 2 punctures with $j=1,2$. For (\ref{newc}) to be
satisfied, $m_2$ cannot be larger than 1 in magnitude, so that
there are only 3 combinations of $m$ possible. As the $j$ values may be
interchanged, there are 6 states in the $j,m$ counting prescription.
However, the $m$ counting prescription yields only 3 because the $j$
values are not to be taken into consideration here. On the other
hand, setting $|m|=j$, as in \cite{agullo}, leads to no state at all
because $\pm 2$ and $\pm 1$ cannot cancel.

In short, in considering small black holes, or any black hole
which can be treated exactly, 
the $|m|=j$ method of \cite{agullo} may not
count all states with distinct $m$-labels. It
gives a severely reduced estimate except in special cases involving $j=\frac12$
or for large area \cite{meissner,gm2}. In general,
to get the correct number of horizon states ($m$ counting), 
one has to use the formula ${(\sum s_m)!\over
\prod s_m!}$, where $s_m\equiv\sum_j s_{j,m}$, \cite{gm2} 
for each allowed set $\{m\}$ and find the sum.

{\em We thank Fernando Barbero for correspondence.}

\end{document}